\documentclass[%
 reprint,
 showpacs,preprintnumbers,
 amsmath,amssymb,
 aps,
prb
]{revtex4-1}
\usepackage{dcolumn}
\usepackage[dvipdfmx]{graphicx}
\usepackage{subfigure}
\usepackage{color}
\usepackage{mathrsfs}

\makeatletter
\def\btt#1{\texttt{\@backslashchar#1}}
\DeclareRobustCommand\bblash{\btt{\@backslashchar}} \makeatother

\def\gsim{\lower -0.3ex \hbox{$>$} \kern -0.75em \lower 0.7ex
	\hbox{$\sim$}}
\def\lsim{\lower -0.3ex \hbox{$<$} \kern -0.75em \lower 0.7ex
	\hbox{$\sim$}}

\begin{document}

 \title{
Spin relaxation in hole-doped transition metal dichalcogenide monolayer and bilayer
 with the crystal defects
 }

 \author{Tetsuro Habe and Mikito Koshino}
 \affiliation{Department of Physics, Tohoku University, Sendai 980-8578, Japan}

 \date{\today}

 \begin{abstract}
 We study the electronic spin relaxation effect in the hole-doped 
monolayer and bilayer transition-metal dichalcogenides in the presence of the crystal defects.
We consider realistic models of the lattice vacancy 
and actually estimate the spin relaxation rate using the multi-orbital tight-binding model.
In the monolayer, the spin-relaxation time is found to be extremely long
compared to the momentum relaxation time,
and this is attributed to the fact that the spin hybridization in the band structure 
is suppressed by the mirror reflection symmetry.
The bilayer TMD has a much shorter spin relaxation time in contrast,
and this is attributed to stronger spin hybridization due to the absence of the mirror symmetry.
 \end{abstract}


 \maketitle


Monolayer transition-metal dichalcogenide(TMD) is a atomically thin two-dimensional semiconductor 
with a strong spin-orbit interaction\cite{Xiao2012}.
In the electronic band structure, the spin and valley ($K$ and $K'$) degrees of freedom are intercorrelated 
because of the broken inversion symmetry in the atomic configuration,
implying  various spin-dependent phenomena and potential applications to the spintronic devices \cite{Xiao2012,Cao2012,Zeng2012,KinFai2012,Shi2013,Molina2013,Suzuki2014,ZhangC2014,Yamamoto2014} and electronics\cite{Song2013,Yuan2014,Feng2012,Klinovaja2013,Kormanyos2014,Peng2014,Habe2015}.

\begin{figure}[htbp]
	\includegraphics[width=90mm]{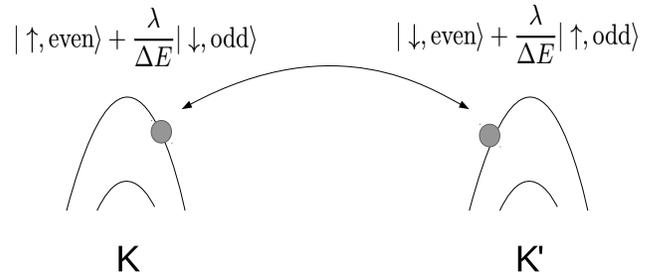}    
	\caption{
	Schematic description of the inter-valley spin scattering in monolayer TMD.
	}\label{inter_valley}
\end{figure}

While the spin-orbit interaction is a key to control the electronic spin,
it also causes the spin relaxation at the same time in presence of the impurity scattering.
In the conventional semiconductors, 
the spin polarization of the conduction electron rapidly decays
in various processes such as Dyakonov-Perel mechanism and Elliott-Yafet mechanism.
\cite{d1971spin,elliott1954theory,kiselev2000progressive,kaneko2008numerical}
In the TMDs, on the other hand, the recent experiment showed that
the carriers in the valence bands have a relatively long spin-relaxation time 
compared to conventional materials \cite{Mak2012}.
The long spin life time comes from the peculiar band structure of TMD,
where the spin-up and spin-down subbands split in opposite direction between $K$ and $K'$ valleys
with a relatively large splitting width of the order of 100 meV [Fig. \ref{inter_valley}].
In the moderate hole concentration, therefore, the carriers at $K$ and $K'$ are fully polarized to opposite spin directions,
and then the inter-valley scattering is necessary for the spin relaxation.

\begin{figure}[htbp]
	\includegraphics[width=85mm]{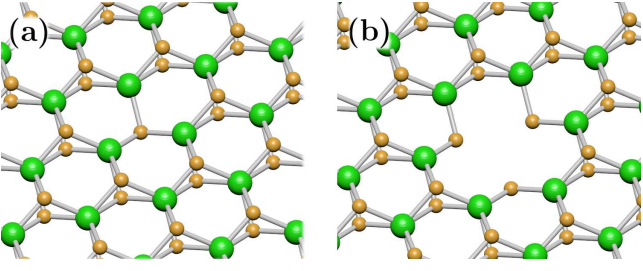}    
	\caption{
	Atomic structure of 
	(a) a single chalcogen-site vacancy ($V_{\rm X}$), and (b)
	multiple vacancies on a transition-metal sites  and three chalcogen sites ($V_{\rm MX_3}$),
	in monolayer MX$_2$.
	}\label{vacancy}
\end{figure}

In this paper, we study the inter-valley spin relaxation effect caused by the non-magnetic short-range scatterers
in  hole-doped monolayer and bilayer TMDs.
In the literature, the spin-relaxation in monolayer TMD was theoretically investigated 
for the conduction / valence bands and various different relaxation mechanisms.
\cite{Wang2014-1, Wang2014-2,Ochoa2013-1,Ochoa2013-2,Ochoa2014}
The spin relaxation by the inter-valley scattering has also been discussed 
using effective impurity models, \cite{Shan2013,Ochoa2014}
while the spin-flip mechanism in actual crystal defect or impurity in the lattice structure
has not been studied well.
In what follows, we consider specific atomic defects \cite{Zhou2013} shown in Fig.\ \ref{vacancy}
which are actually observed in real systems, 
and estimate the spin relaxation rate using the multi-orbital tight-binding model.
There we show that the spin-polarized valence bands at $K$ and $K'$
are actually not pure spin states but includes small components of the opposite spin 
due to the spin-orbit interaction, and this enables the inter-valley scattering 
under the non-magnetic defects.
In the monolayer TMD, such a spin hybridization is relatively weak
because of the limitation by the mirror reflection symmetry $\sigma_h$,
leading to a extremely small spin relaxation rate (long spin relaxation time)
compared to the momentum relaxation rate.
In the bilayer TMD,  on the other hand, 
the spins are more strongly hybridized due to the absence of the mirror reflection symmetry,
and as result,  the spin relaxation rate is shown to be
greater than in monolayer by the factor of $10^3$.

\begin{figure}[htbp]
	\includegraphics[width=60mm]{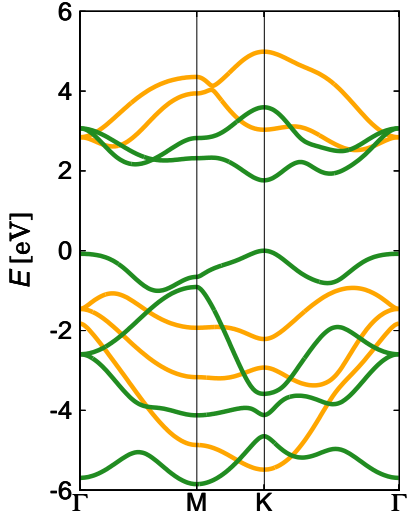}    
	\caption{
	Band structure of monolayer MoS$_2$ without spin-orbit interaction. Even and odd bands are indicated by
	green (dark-colored) and orange (light-colored) curves, respectively.
	}\label{bands}
\end{figure}

In TMDs, the electric states around the Fermi energy consists of the d-orbitals $(d_{3r^2-z^2}, d_{x^2-y^2},d_{xy},d_{xz},d_{yz})$ on transition-metal atoms (M = Mo, W) and the p-orbitals $(p_x,p_y,p_z)$ on chalcogen atoms
(X = S, Se, Te).
As the primitive unit cell contains a single transition-metal atom 
and two chalcogen atoms, we consider eleven atomic orbitals for each spin in a unit cell.
The monolayer TMD has the mirror reflection symmetry $\sigma_h$ with respect to the layer plane, 
and therefore  the atomic orbitals can be classified into even orbitals
 $(d_{3z^2-r^2}, d_{x^2-y^2}, d_{xy}, p_x^+, p_y^+, p_z^-)$, 
 and odd orbitals $(d_{xz},d_{yz}, p_x^-, p_y^-, p_z^+)$,
 with respect to the eigenvalue in inverting the $z$-direction (out-of-plane direction).
Here $p_\mu^\pm=p_\mu^{\mathrm{t}}\pm p_\mu^{\mathrm{b}}$
is the superposition of the atomic orbitals on the vertically located pair of two chalcogen $p$-orbitals
with $p_\mu^{\rm t}$ and $p_\mu^{\rm b}$ for top and bottom.
In the absence of the spin-orbit interaction, the even and odd orbitals
independently form the even and odd energy bands, respectively.
Fig.\ \ref{bands} shows the band structure of MoS$_2$
calculated by the first-principle method of quantum-ESSPRESSO\cite{Quantum-espresso}.

To describe the motion of electrons, we derive the tight-binding model 
from the first principle band structure
by using Wannier90, which is a numerical package for minimized wannnier functions\cite{Wannier90}. 
We first create the tight-binding Hamiltonian disregarding the spin-orbit interaction,
and then introduce $\boldsymbol{L}\cdot\boldsymbol{s}$ spin-orbit term to the transition metal atoms.
The Hamiltonian for the spin-independent part is written as
\begin{align}
H_0=\sum_{i,j,s}{\boldsymbol{a}^{s}_i}^{\dagger}
t^{\rm e}({i,j})
\boldsymbol{a}^{s}_{j}
+
{\boldsymbol{b}^{s}_i}^{\dagger}
t^{\rm o}({i,j})
\boldsymbol{b}^{s}_{j}
\end{align}
where $\boldsymbol{a}^s_{i}$ and $\boldsymbol{b}^s_{i}$ are the annihilation operators 
for the six even orbitals and the five odd orbitals in the unit cell $i$, respectively, and with $s=\pm$
 represent the spin degree of freedom.
The spin-orbit interaction for the $d$-orbitals of transition-metal atom is described 
in the basis of ($d_{3z^2-r^2}$, $d_{x^2-y^2}$, $d_{xy}$, $d_{xz}$, $d_{yz}$) as \cite{Condon1935,Shanavas2014} 
\begin{align}
\tilde{H}_{\mathrm{SO}}=&
\frac{i\lambda}{2}
\left(
\begin{array}{ccc|cc}
0&0&0&-\sqrt{3}\sigma_y&\sqrt{3}\sigma_x\\
0&0&-2\sigma_z&\sigma_y&\sigma_x\\
0&2\sigma_z&0&-\sigma_x&\sigma_y\\
\hline
\sqrt{3}\sigma_y&-\sigma_y&\sigma_x&0&-\sigma_z\\
-\sqrt{3}\sigma_x&-\sigma_x&-\sigma_y&\sigma_z&0
\end{array}
\right),
\label{spin_orbit}
\end{align}
Here the first three orbitals ($d_{3z^2-r^2}$, $d_{x^2-y^2}$, $d_{xy}$) are even 
and the latter two ($d_{xz}$, $d_{yz}$) are odd under the mirror reflection $\sigma_h$.
Since the spin flipping terms ($\sigma_x$ and $\sigma_y$)
always appear in the off diagonal block connecting even and odd orbitals,
the spin orbit Hamiltonian can be written by
\begin{align}
H_{\mathrm{SO}}=&\sum_{s,i}
{\boldsymbol{a}^{s}_i}^{\dagger}
{u^{\mathrm{e}}}({s}) 
\boldsymbol{a}^{s}_i
+
{\boldsymbol{b}^{s}_i}^{\dagger}
{u^{\mathrm{o}}}({s}) 
\boldsymbol{b}^{s}_i
\nonumber\\
&+\left[{\boldsymbol{a}^{s}_{\boldsymbol{u}_i}}^{\dagger}u^{\mathrm{off}}(s,\bar{s})\boldsymbol{b}^{\bar{s}}_{\boldsymbol{u}_j}+\mathrm{h.c}\right],
\end{align}
where $\bar{s}$ represents the opposite spin to $s$,
$u^{\rm e}$ and $u^{\rm o}$ represent the diagonal blocks for even and odd states, respectively,
and $u^{\rm off}$ is the off-diagonal block.
We neglect the spin-orbit interaction in the chalcogenide atom which is much smaller than that in the transition-metal atom.


Before the detailed numerical calculation, 
we consider the mechanism of the spin-relaxation through the inter-valley scattering 
in the hole-doped TMD monolayer.
It is similar to the conventional Elliot-Yafet process,
while the even-odd classification due to the reflection symmetry
imposes a limitation to possible scattering process.
In TMD, the low-energy spectrum near the band gap is dominated by the even-orbital bands.
The diagonal part of $H_{\rm SO}$  is responsible for the band splitting 
between spin-up and spin-down branches due to $\pm \sigma_z$ terms \cite{Xiao2012}.
As schematically shown in Fig.\ \ref{inter_valley},
the valence bands are spin-split in opposite directions between two valleys
due to the time-reversal symmetry,
and in the moderate hole-doped regime considered in the following,
the Fermi energy crosses only $(K, \uparrow)$ and $(K', \downarrow)$ branches.

\begin{figure}[htbp]
	\includegraphics[width=80mm]{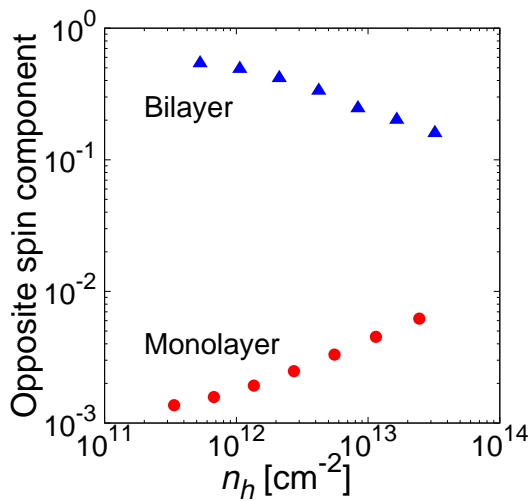}    
	\caption{
	Relative amplitude of spin-down component included in the up-spin state
        in MoS$_2$ monolayer and bilayer,
       as function of the hole density $n_h$.
	}\label{spin_component}
\end{figure}

The off-diagonal part of $H_{\rm SO}$  hybridizes these low-energy even-state bands with
the odd-state bands far from the Fermi energy with the opposite spin.
The states near $K$ and $K'$ points are then expressed in the first-order perturbation as
\begin{align}
& |K, \uparrow \rangle \approx  
|\mathrm{even}, \uparrow\rangle+\frac{\Lambda}{\Delta E}|\mathrm{odd}, \downarrow\rangle
\nonumber\\
&|K', \downarrow \rangle \approx  
|\mathrm{even}, \downarrow\rangle+\frac{\Lambda}{\Delta E}|\mathrm{odd}, \uparrow\rangle,
\label{eq_K_states}
\end{align}
where  $|\mathrm{even (odd)}, s \rangle$  represents the eigenstate
in the absence of spin-orbit coupling, which is 
a direct product of the even (odd) orbital state and the pure spin state with $s$.
Here $\Lambda$ is the energy scale of the coupling matrix elements,
and $\Delta E$ is the typical energy distance from the Fermi energy to the odd state bands.
We can show that $\Lambda$ vanishes right at $K$ and $K'$ points \cite{Ochoa2014},
and linearly increases as the wave number is shifted from the valley center.
Apart from the numerical factor (of the order of 1), it is roughly written as 
\begin{align}
\Lambda \sim \lambda k a,
\label{eq_Lambda}
\end{align}
where $\lambda$ is the spin-orbit coupling constant in Eq.\ (\ref{spin_orbit}),
$k$ is the relative wave-vector from $K$ or $K'$ (omitted in Eq.\ (\ref{eq_K_states})),
and $a$ is the atomic scale constant.
Fig.\ \ref{spin_component} shows the relative amplitude
of spin-down component included in the up-spin state
in MoS$_2$ monolayer (indicated by red dots),
which is averaged over the Fermi surface and 
plotted against the corresponding hole density $n_h$. 
We see the amplitude is roughly proportional to $\sqrt{n_h} \propto k$
in consistent with Eq.\ (\ref{eq_Lambda}).

We assume the scatterer $H_i$ is non-magnetic and does not flip spins.
Then the scattering matrix element between $|K, \uparrow \rangle$
and $|K', \downarrow \rangle$ is written as
\begin{align}
\langle K', \downarrow|H_i|K, \uparrow\rangle
=\frac{\Lambda}{\Delta E}
&[
\langle 
\mathrm{odd}, \uparrow
|H_i|
\mathrm{even},\uparrow\
\rangle
\nonumber\\
&+
\langle
\mathrm{even}, \downarrow
|H_i|
\mathrm{odd},\downarrow\
\rangle
].
\end{align}
We see that $H_i$ is required to break the mirror reflection symmetry
to hybridize the odd and even orbital states.
If we consider an atomic-scale scattering potential breaking mirror symmetry,
we roughly have $\langle \mathrm{even},s |H_i| \mathrm{odd}, s\rangle  \sim  V a^2/S $,
where  $V$ is the energy scale of the potential
and $S$ is the total system area.
As a result,  the inter-valley matrix element can be estimated as
 \begin{align}
\langle K' \downarrow|H_i|K \uparrow\rangle \sim 
V \frac{\Lambda}{\Delta E}
\frac{a^2}{S}
\end{align}
When $V$ is small and the scatterers are distributed sparsely in the system
with a number density $n_i$, the inter-valley spin relaxation rate (inverse spin relaxation time) is
\begin{align}
\Gamma_{\rm s} 
\sim
2\pi
|\langle K' \downarrow|H_i|K \uparrow\rangle|^2 n_i\rho S^2,
\end{align} 
where $\rho$ is the density of states at the Fermi energy per valley and per area.

On the other hand, the momentum scattering rate 
is dominated by the intra-valley scattering process without spin flip, and it is written as
\begin{align}
\Gamma_{\rm p} 
\sim 
2\pi
|\langle K \uparrow|H_i|K \uparrow\rangle|^2 n_i\rho S^2,
\end{align}
where $\langle K \uparrow|H_i|K \uparrow\rangle \sim V a^2 /S$.
The ratio of two scattering rates then becomes
\begin{align}
\frac{\Gamma_{\rm s}}{\Gamma_{\rm p}}
\sim
\left(
\frac{\Lambda}{\Delta E}
\right)^2
=
2\pi n_h a^2
\frac{\lambda^2}{(\Delta E)^2},
\label{eq_gs_gp}
\end{align} 
where we used $\Lambda \sim \lambda k_F a$, and 
$n_h= 2 \pi k_F^2 / (2\pi)^2$ is the number of holes per unit area.
The linear dependence of $\Gamma_{\rm s}/\Gamma_{\rm p}$ on the hole-density $n_h$ originates from
the $k$-dependence of $\Lambda$.

For MoS$_2$, the strength of the spin-orbit interaction $\lambda$ is given by
$\lambda\approx 0.073$ eV\cite{Liu2013},
and the odd-even energy distance $\Delta E$ is typically 2 eV.
The lattice constant is $a\sim 0.3$ nm.
At the moderate hole density of $n_h = 10^{12}$ cm$^{-2}$, 
for example, the ratio $\Gamma_{\rm s}/\Gamma_{\rm p}$ becomes $10^{-5}$.
When assuming the disorder scattering rate $\Gamma_{\rm p} \sim 0.01$ eV, for example,
we have the spin-relaxation time can be estimated to $\tau_s=\hbar/\Gamma_{\rm s} \sim 10$ ns.

\begin{figure}[htbp]
	\includegraphics[width=85mm]{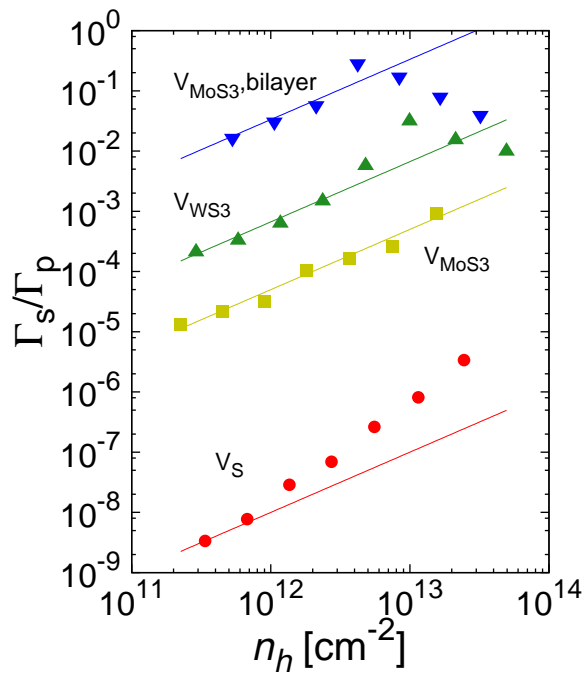}
	\caption{
	Relative spin-relaxation rate  $\Gamma_{\rm s}/\Gamma_{\rm p}$
	as a function of hole density of $n_h$, calculated
	 for $V_{\rm S}$ and $V_{\rm MoS_3}$ in the monolayer MoS$_2$,
$V_{\rm WS_3}$ in the monolayer WS$_2$.
and $V_{\rm MoS_3}$ in the bilayer MoS$_2$.
Solid lines just represent the linear dependence to $n_h$.
	}\label{relaxation_time}
\end{figure}


The realistic short-ranged scatterers such as atomic defects or vacancies cannot be treated
as a perturbational potential, and then
we need to actually solve the scattering problem to estimate the spin relaxation rate.
We consider two specific mirror-symmetry-breaking configurations 
which are actually observed in the real materials\cite{Zhou2013}:
a single chalcogen-site vacancy ($V_{\rm X}$),
and a multiple vacancy on a transition-metal sites 
and three chalcogen sites ($V_{\rm MX_3}$),
which are shown in Fig.\ \ref{vacancy} (a) and (b), respectively.
We consider a system with a single defect of each type in a tube geometry with a sufficiently large diameter,
and calculate the transfer matrix $T$  using the tight-binding model with
the one-dimensional transport formula \cite{Ando1991}.

The scattering rate from the initial state $(s,\boldsymbol{k})$ is given 
by the $T$ matrix as,
\begin{align}
\frac{1}{\tau^\pm_{s\boldsymbol{k}}}
=
\frac{2\pi n_i{S}}{\hbar}
\sum_{s'\boldsymbol{k}'}
\left|T_{s\boldsymbol{k},s'\boldsymbol{k}'}\right|^2 
\delta(\varepsilon_{s\boldsymbol{k}}-\varepsilon_{s'\boldsymbol{k}'})\, \delta_{s',\pm s},
\label{eq_tau_def}
\end{align}
where $S$ is the system area,
and $\tau^+$ describes the spin-conserving scattering time
and $\tau^-$ the spin-flipping scattering time.
The momentum relaxation rate and the spin relaxation rate are obtained by
\begin{align}
& \Gamma_{\rm p} = \left\langle \frac{1}{\tau^+_{s\boldsymbol{k}}} + \frac{1}{\tau^-_{s\boldsymbol{k}}}\right\rangle_{\varepsilon_F} \nonumber\\
& \Gamma_{\rm s} = \left\langle \frac{1}{\tau^-_{s\boldsymbol{k}}}\right\rangle_{\varepsilon_F},
\end{align}
respectively, where $\langle\cdots \rangle_{\varepsilon_F}$ represents
the average on the Fermi surface.

First we present the analyses for $V_{\rm S}$ and $V_{\rm MoS_3}$ in the monolayer MoS$_2$,
and also for $V_{\rm WS_3}$ in the monolayer WS$_2$.
Fig.\ \ref{relaxation_time} plots the relative spin-relaxation rate $\Gamma_{\rm s}/\Gamma_{\rm p}$ 
averaged on the Fermi surface, as a function of the corresponding hole density $n_h$.
Here $\Gamma_{\rm p}$ is found to be
always about $n_i a^2$ times a few eV not strongly depending on $n_h$,
so it is reasonable to normalize the spin relaxation rate by $\Gamma_{\rm p}$.
For  $V_{\rm MoS_3}$ in the monolayer MoS$_2$,
we see that the $\Gamma_{\rm s}/\Gamma_{\rm p}$ 
is roughly proportional to $n_h$ in the low density regime,
and this is consistent with the qualitative estimation Eq.\ (\ref{eq_gs_gp}).
The absolute value is of the order of $10^{-5}$ at $n_h \sim 10^{12}$ cm$^{-2}$,
which is also in a good agreement with the estimation.
This is somewhat surprising because Eq.\ (\ref{eq_gs_gp})
was derived in the perturbational approach which is valid in the weak potential limit.

$V_{\rm WS_3}$ in the monolayer WS$_2$ exhibits a similar behavior,
while the absolute value is greater than in MoS$_2$ about by factor of 10.
WS$_2$ has a larger spin-orbit interaction of
$\lambda_{\mathrm{WS}_2} \approx0.211\mathrm{eV}$.\cite{Liu2013}
In Eq.\ (\ref{eq_gs_gp}), the relative spin-relaxation rate $\Gamma_{\rm s}/\Gamma_{\rm p}$ 
is proportional to $\lambda^{2}$, and the order-of-magnitude difference is actually consistent with 
$(\lambda_{\mathrm{WS}_2}/\lambda_{\mathrm{MoS}_2})^{2}\approx8.35$.
For $V_{\rm WS_3}$, we see some non-monotonic behavior in higher $n_h$.
This may be related to the impurity bound states, while the detailed study 
is out of the scope of the present work.

In a simpler defect $V_{\rm S}$, the spin relaxation rate is found to be 
extremely small compared to that of $V_{\rm MoS_3}$,
and it is also deviating from a linear function of $n_h$.
In the low-energy bands, the wave amplitude of the electronic state 
mainly resides on the transition-metal atoms, and thus a single chalcogen vacancy
is expected to have relatively small effect compared to $V_{\rm MoS_3}$.
But it seems not enough to fully explain the difference of the order of a few magnitudes,
and we presume some other factor, e.g., cancellation of the leading-order matrix elements,
may also contribute to the suppression of the intervalley process.

Finally, we study the spin relaxation rate in the MoS$_2$ bilayer with a $V_{\rm MoS_3}$ defect.
Unlike monolayer, the energy bands in the bilayer TMD are all spin degenerate 
due to the coexistence of time reversal symmetry and spatial inversion symmetry.\cite{Xiao2012}
However, each eigenstate cannot be expressed as a single spin state
because the spin-orbit interaction strongly hybridizes the different spin components in the multiple orbitals.
To define the spin relaxation rate using Eq.\ (\ref{eq_tau_def}) in this situation, 
we reconstruct the degenerate Bloch states at each Bloch momentum
into approximate "spin-up" and "spin-down" states so as to maximize the expectation value of $\sigma_z$.
The relative amplitude of the spin-down components in the "spin-up" state in bilayer MoS$_2$
is shown in Fig.\ \ref{spin_component}, where we still see a strong hybridization about a few of 0.1.
In the monolayer, in contrast, the spin hybridization is extremely weak as we have already seen,
and this is because the selection rule imposed by the mirror reflection symmetry
allows the spin mixing only between the low-energy even bands and the high-energy odd bands.
In bilayer, the mirror reflection symmetry is absent 
and there the spin-orbit interaction directly couples the opposite spin states in the same low-energy bands,
giving a strong spin hybridization.
The spin relaxation rate $\Gamma_{\rm s}/\Gamma_{\rm p}$ 
 calculated for bilayer MoS$_2$ is plotted in Fig.\ \ref{relaxation_time}, 
and it is much greater than that of monolayer by the factor of $10^3$
reflecting the strong spin mixing effect.
 
 To conclude, we studied the electronic spin relaxation 
in hole-doped TMD monolayer and bilayer in the presence of realistic atomic defects.
By analyzing the band structure and spin-orbit interaction,
we  qualitatively describe the spin relaxation mechanism
and actually estimated the spin-relaxation rate for several specific cases in the numerical calculations.
In the monolayer TMD, the inter-band spin hybridization 
is suppressed by the parity-selection rule under the mirror symmetry,
resulting in a relatively smaller spin relaxation rate.
The bilayer TMD has a much greater spin relaxation rate in contrast,
because of the strong spin hybridization in the absence of the mirror symmetry.

This work was supported by Grants-in-Aid for Scientific research (Grants No.\ 25107005).

\bibliography{TMD}

\end{document}